\documentstyle[12pt,epsf]{article}

\newcommand {\be}{\begin{equation}}
\newcommand {\ee}{\end{equation}}
\newcommand {\x}{{\mathbf x}}
\newcommand {\xxi}{{\mathbf \xi}}
\newcommand {\msbox}[1]{\mbox{\scriptsize #1}}

\begin{document}
\title{Monte Carlo Simulations of Off-Lattice Polymers}
\author{Peter Grassberger and Rainer Hegger\\
Physics Department, University of Wuppertal\\
42097 Wuppertal, Germany}
\maketitle

\begin{abstract}
We point out that a newly introduced recursive algorithm for lattice 
polymers has a much wider range of applicability. In particular, we 
apply it to the simulation of off-lattice polymers with Lennard-Jones 
potentials between non-bonded monomers and either delta or harmonic 
potentials between bonded monomers. Our algorithm allows particularly 
easy calculations of the free energy, and seems in general more efficient 
than other existing algorithms.  
\end{abstract}
\newpage

\section{Introduction}

In a number of recent papers, new Monte Carlo schemes for simulating 
off-lattice polymers have been proposed \cite{kumar-sl}--\cite{sheng}. 
A particular aspect there was the calculation of chemical potentials.
This is not easy in most schemes, which has lead to some controversy 
\cite{smit,kumar2}. In this note we want to point out that a recently 
introduced recursive implementation of the enrichment method for lattice 
polymers \cite{gh} can be easily adapted to this problem. It is both easy 
to implement and efficient, and the computation of the chemical potential 
is straightforward. Indeed, for polymers without long-range 
monomer-monomer interactions and without interactions with any solvent 
our method seems to be faster than all methods mentioned above. It is 
{\it not} efficient for systems at extremely low energies 
(when Boltzmann factors due to attractive potentials between individual 
monomers become $> 10$), and for systems with long-range forces 
\cite{joensson,irbaeck}. 

Though we shall apply our algorithm only to polymer systems, we should point 
that it is much more general. It can be applied to any equilibrium system 
which can be broken up into discrete units, labeled by an index 
$i=1,\ldots N$, and whose internal energy can be written as 
\be
    U = \sum_{i=1}^N U_i\,,
\ee
where $U_i$ depends {\it only} on units with label $i'\leq i$. In the case 
of a polymer with potential $u_{ij}$ between non-bonded monomers and 
potential $v_i$ between monomers $i$ and $i-1$, we choose of course 
$U_i = v_i + \sum_{j<i} u_{ij}$ for $i>1$, while $U_1=v_1$ is evaluated 
with $\x_0=0$. Thus the start of polymer chain is anchored to 
$x=0$.

The algorithm basically tries to build the system by assembling it unit 
by unit. To obtain the correct Boltzmann weights, the entire assembled 
configuration has to be discarded occasionally (with probabilities 
depending on these weights) if the following units do not fit. A typical 
example is a self-avoiding walk on a lattice, where the configuration 
has to be discarded if the next step would lead to a self intersection. 
In order to compensate for this ``attrition", we use an ``enrichment" 
method, the basic idea of which was already introduced more than 30 years 
ago \cite{wall}. Instead of trying just one continuation from a partially 
assembled system, the intermediate sample is ``enriched" by replicas 
which serve as starting points for independent continuations. Thus from 
each partly assembled system more than one continuation is attempted. 
These attempts are made irrespective of whether one of them is successful 
or not, which leads immediately to an unbiased Gibbs ensemble and which 
distinguished the method from most other proposals to overcome attrition.

But in contrast to earlier implementations of the enrichment idea which 
were of ``breadth first" type \cite{grishman,garel,higgs,velikson}, our 
implementation is ``depth first" \cite{tarjan}. This implies completely 
different data structures. In particular, it means that a partly assembled 
system can reside in the fast memory of the computer even in very large 
simulations. This avoids the very time consuming data transfer needed 
in breadth-first implementations, unless the systems are very small. As 
discussed in \cite{higgs}, simulating a system with $N$ units one has to
simulate $>>N$ ensemble members simultaneously in a breadth first algorithm. 
This requires a storage space $> {\cal O}(N^2)$, which limits severely 
the system sizes. But even if the entire ensemble fits into memory, a 
breadth first algorithm is slower by a factor ${\cal O}(N)$ since adding 
a new unit takes a time ${\cal O}(N)$ \cite{higgs} (there is a finite 
probability that a new replica has to be created) while it needs only a 
time ${\cal O}(1)$ for a depth first algorithm. Also, the simplest and 
most intuitive implementation of a depth first algorithm is via recursive 
function calls. In this case the compiler makes all the book-keeping 
which is fast but quite non-trivial in this approach. Thus our method 
avoids all problems which have made the enrichment method unpopular in 
the past.

The only disadvantage of a depth first implementation is that we have to 
guess the attrition in advance. In a breadth-first approach, level $i$ is 
treated only after all previous levels have been finished, and thus the 
attrition on the previous levels is known. If we know the amount of attrition 
sufficiently well from other sources, we do not have any problem in a depth 
first approach either. Otherwise, the best strategy is to start with small 
systems and a conservative estimate of the attrition, and to increase the 
system size in separate runs as the attrition gets better and better known 
\cite{gr-disorder}.

\section{Algorithm} 

Our aim is to compute the partition function 
\be
   Z_N = \int d\x_1\ldots d\x_N \,e^{-\beta U(\x_1\ldots\x_N)} \;,
      \qquad \beta=(k_BT)^{-1} \;.
\ee
In addition, we consider also ``partial partition functions" which 
describe the last $N-i$ units in the static background field created 
by the first $i+1$ units, 
\be
   Z_{N-i|i}(\x_1\ldots\x_i)=\int d\x_{i+1}\ldots d\x_N \;
           e^{-\beta\sum_{j=i+1}^N U_j(\x_1\ldots\x_j)} \;.
\ee
They can be written recursively, 
\be
   Z_{N-i+1|i-1}(\x_1,\ldots,\x_{i-1})= \int d\x_i \;e^{-\beta 
     U_i(\x_1\ldots\x_i)} \; Z_{N-i|i}(\x_1,\ldots,\x_i) \label{part-recurs}
\ee
with $Z_{0|N}(\x_1,\ldots,\x_N)\equiv 1$ and
$Z_{N|0}\equiv Z_N$. The basic strategy will be to compute Monte
Carlo estimates for the partial partition functions using this
recursion relation (this will be done implicitly, and the explicit
code needed to do it will be very compact). The total partition
function is generated by ``assembling" the last units first (which
means just summing over suitable statistical samples), and working
one's way back. The task is completed when finally the sample points
for the first unit are summed over. With this strategy in mind we will 
in the following concentrate on one typical step in the recursion
relation where $Z_{N-i|i}(\x_1\ldots\x_i)$ is assumed to be known, and
$Z_{N-i+1|i-1}(\x_1\ldots\x_{i-1})$ is to be computed.

We assume that the potential $U_i$ can be split into two parts, 
\be
   U_i(\x_1\ldots \x_i) = U_i^{(0)}(\x_1\ldots \x_i) + 
                \Delta U_i(\x_1\ldots \x_i) \;,
\ee
with the following properties: \\

(i) The partial partition functions $Z_{N-i|i}^{(0)}$ associated with 
$U^{(0)}$ can be computed for each background configuration $(\x_1\ldots
\x_i)$ either analytically or by some other method which is independent 
of the present Monte Carlo method;

(ii) A fast (pseudo-)random number generator exists which produces points 
$\xxi_k$ distributed with the density 
\be
   \rho^{(0)}_i(\xxi|\x_1\ldots \x_{i-1}) = 
           {Z_{N-i|i}^{(0)}(\x_1\ldots\x_{i-1},\xxi)\over 
            Z_{N-i+1|i-1}^{(0)}(\x_1\ldots\x_{i-1})}\, 
           e^{-\beta U_i^{(0)}(\x_1\ldots\x_{i-1},\xxi)}  \;
\ee
(notice that this is normalized due to eq.(\ref{part-recurs}));

(iii) $\Delta U_i$ is ``kind'' in the sense that the integral 
$\int d\xi\rho^{(0)}_i e^{-\beta \Delta U_i}$ converges for all 
background configurations and is not too big. We should stress 
that the last condition affects only the efficiency of the method, 
but is not related to any bias. In particular, we make no series 
expansion or truncation in higher powers of $\Delta U_i$, or anything 
of that sort. In general, it will be sufficient if $U_i^{(0)}$ has 
correct asymptotic behavior for the integral to converge, and has 
roughly the same shape as $U_i$. 

Our $U^{(0)}$ is similar to the ``guiding field" in \cite{garel,higgs}.

Using this decomposition of $U_i$ one shows easily that 
\begin{eqnarray}
   Z_{N-i+1|i-1}(\x_1\ldots\x_{i-1})  & = & Z_{N-i+1|i-1}^{(0)}
          (\x_1\ldots\x_{i-1})
          \int d\xxi \,\rho^{(0)}_i(\xxi|\x_1\ldots \x_{i-1}) \; \\
       &  & e^{-\beta \Delta U_i(\x_1\ldots\x_{i-1},\xxi)} \,
            {Z_{N-i|i}(\x_1\ldots\x_{i-1},\xxi) \over 
             Z_{N-i|i}^{(0)}(\x_1\ldots\x_{i-1},\xxi)}  \;.
\end{eqnarray}
The integral over $\xxi$ can now be approximated by a sum over random 
points $\xxi_k$ obtained by means of the above random number generator, 
and we obtain
\begin{eqnarray}
   Z_{N-i+1|i-1}(\x_1\ldots\x_{i-1}) & = & Z_{N-i+1|i-1}^{(0)}
          \; \lim_{M\to\infty} {1\over M} \sum_{k=1}^M         \\
        &   & e^{-\beta \Delta U_i(\x_1\ldots\x_{i-1},\xxi_k)} \;
           {Z_{N-i|i}(\x_1\ldots\x_{i-1},\xxi_k) \over 
             Z_{N-i|i}^{(0)}(\x_1\ldots\x_{i-1},\xxi_k)}  \;.
\end{eqnarray}
Assume that we have already an estimator for $Z_{N-i|i}$. Then an 
estimator for $Z_{N-i+1|i-1}$ is obtained by either associating a weight 
\be
   w_i(\xxi_k;\x_1\ldots\x_{i-1}) \;=\; {1\over M}\,
       {Z_{N-i+1|i-1}^{(0)}(\x_1\ldots\x_{i-1})
        \over Z_{N-i|i}^{(0)}(\x_1\ldots\x_{i-1},\xxi_k)} \;
         e^{-\beta \Delta U_i(\x_1\ldots\x_i,\xxi_k)}
\ee
with each $\xxi_k$, or --- and this is the method used in our approach --- 
by replacing each $\xxi_k$ in the average by $p_i w_i$ replicas of itself, 
each counted with weight 1 and labeled by an index $\alpha$. Here $p_i$ 
is a parameter (independent of $\x_i$ and $\xxi_k$) which is in principle 
arbitrary (more about it will be said 
below) and which controls the size of the sample by counterbalancing the 
attrition\footnote{We use the word ``attrition" for conformity with the 
use in the literature on self-avoiding walks. It should be noted that in 
our case it does not necessarily imply a depletion of chains, but can 
also imply the opposite, depending on the sign of the potential.} during 
the step $i\to i-1$. This gives then our MC estimate
\be
    Z_{N-i+1|i-1}(\x_1\ldots\x_{i-1}) \approx {1\over p_i}\sum_{k=1}^M 
           \sum_{\mbox{\tiny replicas}\,\alpha} 
           Z_{N-i|i}^{[\alpha]}(\x_1\ldots\x_{i-1},\xxi_k) \;.
                            \label{ZZZ}   
\ee
The superscript $\alpha$ on the partial partition function on the rhs. 
is to indicate that we use of course different sample points $(\x_{i+1}
\ldots\x_N)$ for each replica, even though the backgrounds are the same. 

The parameter $p_i$ 
has to be chosen carefully: large $p_i$ will lead to samples whose sizes 
increase quickly with $i$, leading to excessive CPU times for large $N$, 
while small $p_i$ lead to too small samples for large $i$. Optimally, 
$p_i$ should be chosen such that the sample size is roughly independent 
of $i$. Notice that this means in particular that $M$ will not be large. 
Indeed, most numerical results quoted below are obtained with $M=1$. 
Large statistics is not obtained by trying many different continuations 
of each partially built chain, but by making many independent runs. 
It is only for extremely low temperatures (not studied here) that $M>>1$ 
should be advantageous since it allows a more uniform covering of 
configuration space. 

We should point out that the above two possibilities (of using $w_i$ 
either as a weight or as a multiplicity of replicas) are indeed just two 
special cases within a much wider range of possible choices. They are all 
distinguished by a different balance between equidistribution of the 
statistical sample and equal weights put on all sample point. It might 
well be that for different purposes different variants are optimal, but 
we shall in the following discuss only the choice of uniform weights, 
corresponding to perfect importance sampling. It has the advantage that 
all thermal averages are just normal averages without any additional 
weight factors except for the weights $p_i$. In particular, the 
incremental chemical potential is simply 
\be
    \mu_i = -k_BT \log{Z_i\over Z_{i-1}}
         \approx - k_BT \log{m_i\over m_{i-1}p_i}\;
\ee
where $m_i$ is the total number of sample point replicas at level $i$ 
(i.e., the total number of subroutine calls at depth $i$ in the algorithm 
described below).

Technically, our method is implemented by means of a recursively called 
subroutine which has the level $i$ as argument. Basically, it just 
chooses a random point $\xxi$, inserts a monomer at this point and computes 
its weight $w_i(\xxi)$, and makes in the average $p_iw_i(\xxi)$ calls to 
itself if $i<N$, with new argument $i+1$. After returning from all these 
subroutine calls, the monomer at $\xxi$ is removed and the subroutine 
is left. This is of course complemented 
by updating the statistics for whatever observable is to be measured. 

Finally, we have to specify what we mean by ``make ... calls {\it in 
average}". In principle, we can choose any distribution for the number 
of calls, provided it gives the right average. But efficiency will depend 
on this choice. One possibility would be e.g. a Poissonian (this would be 
in spirit with the first application of this method in a lattice model 
\cite{gr-disorder}), 
but in general this is not the best choice. As in lattice models, it seems 
that the optimal choice depends on the specific situation, and in extreme 
cases (steep potentials and low temperatures) some experimentation will be 
needed. As a rule of thumb we propose to choose the distribution such that 
it has the smallest possible variance \cite{gh}. Thus, if $p_iw_i(\xi)=k
+ \eta$ with $\eta\in[0,1)$ and $k$ integer, we first make $k$ calls and 
then add one more call with probability $\eta$.

\section{Applications}

We applied this to two versions of a model where non-bonded monomers 
interact by Lennard-Jones potentials in 3-d space, 
\be
   u_{ij} = 4[r_{ij}^{-12}-r_{ij}^{-6}]\;,\quad r_{ij}=|{\mathbf r}_{ij}|
        = |\x_i-\x_j|\;.
\ee
In the first version \cite{kumar}, the force between bonded monomers is 
harmonic, 
\be
   v_i = {\kappa\over 2} \;(r_{i,i-1}-1)^2\;,\quad \kappa=400 \label{harmon} 
\ee
for $r_{i,i-1}>0.5$. For $r_{i,i-1} <0.5$, the potential is infinite. 
In the second version \cite{frenkel}, it is simply provided by hard 
rods which keep a fixed distance $r_{i,i-1}=1$. Notice that we did not 
truncate the potential at large distances, in contrast to previous 
studies \cite{kumar-sl,frenkel}. As far as we can see, the approximations 
involved in correcting for this truncation should be the only possible 
source for eventual disagreements with these works. 

In the second version, a natural choice of $U_i^{(0)}$ is given by an 
isotropic delta potential\footnote{By this we mean of course not strictly 
a delta function but a potential which is enough singular to give a 
delta function for the equilibrium density, $e^{-\beta U_i^{(0)}}=
\delta(r_{i,i-1}-1)$.} fixing $r_{i,i-1}$ at 1. Thus the vectors $\xi$ 
were chosen randomly on the surface of a sphere centered at $x_{i-1}$. 
The corresponding partial partition functions are $Z_{n|i}^{(0)}
=Z_n^{(0)}=(4\pi)^n$.

In the first version the 
choice is less obvious since it is not so easy to produce random points 
according to $e^{-\beta v_i}$ with $v_i$ given by eq.(\ref{harmon}). We 
thus took 
\be
   U_i^{(0)} = v_i(r_{i,i-1})+k_BT\ln(r_{i,i-1}^2)
\ee
For this choice the radial distribution function is a Gaussian centered 
at $r=1$, $\rho^{(0)}({\mathbf r}_{i,i-1})\propto r_{i,i-1}^{-2} 
e^{-\beta v_i}$, and $Z_n^{(0)}=[(2\pi)^3/(\kappa\beta)]^{n/2}$ for 
$\kappa\beta>>1$~\footnote{If this condition is not fulfilled we get
an additional term proportional to an error function due to the
centering of $r_{i,i-1}$ around 1.}.

\begin{figure}[htb]
\centerline{\epsfbox{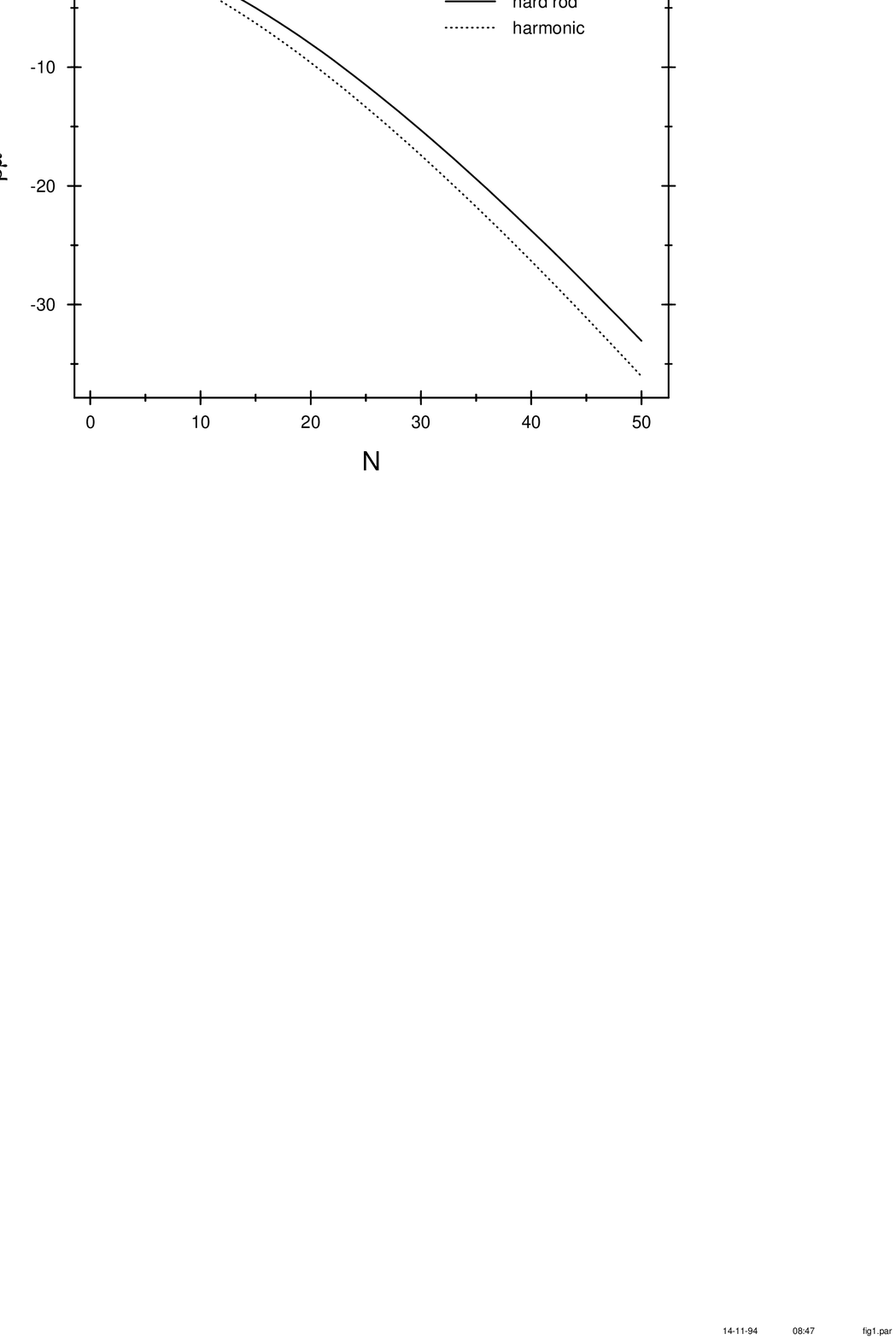}}
\caption{\small Results for the excess chemical potential at $T=1/\beta=1.2$. 
The upper curve is related to the hard rod potential, while the lower
one is the result for the harmonic potential of
eq.(\protect{\ref{harmon}}).  Statistical errors are smaller than the
thickness of the line.}
\label{fig.mu}
\end{figure}

Fig.\ref{fig.mu} shows our data for the excess chemical potential
$\beta\mu^{\mathrm{ex}}$ for the hard rod and the harmonic potential,
respectively. The value $\beta=1/1.2$ was chosen to compare our data
to those of \cite{frenkel}.  From fig.5 of that paper we see that
$\beta\mu^{\mathrm{ex}} = -10.4\pm 0.4$ for a hard rod chain with
$N=30$ monomers. This is much smaller (in absolute value) than our
estimate$\beta\mu^{\msbox{ex}}=-15.35\pm0.05$. Part of this
discrepancy can be explained by the fact that the Lennard-Jones
potential was truncated in \cite{frenkel} at $r=r_c\equiv 2.5$ and its
contribution from $r>r_c$ was estimated analytically. This was not
done in our simulations, where all potentials were taken into account
exactly. But by performing the same truncation as in \cite{frenkel}
without correcting for it at all, we estimated that even this is much
too small ($\approx 9-10\%$) to explain the discrepancy. The
simulations of \cite{kumar-sl} have too large statistical errors for a
similarly detailed comparison.

\begin{figure}[htb]
\centerline{\epsfbox{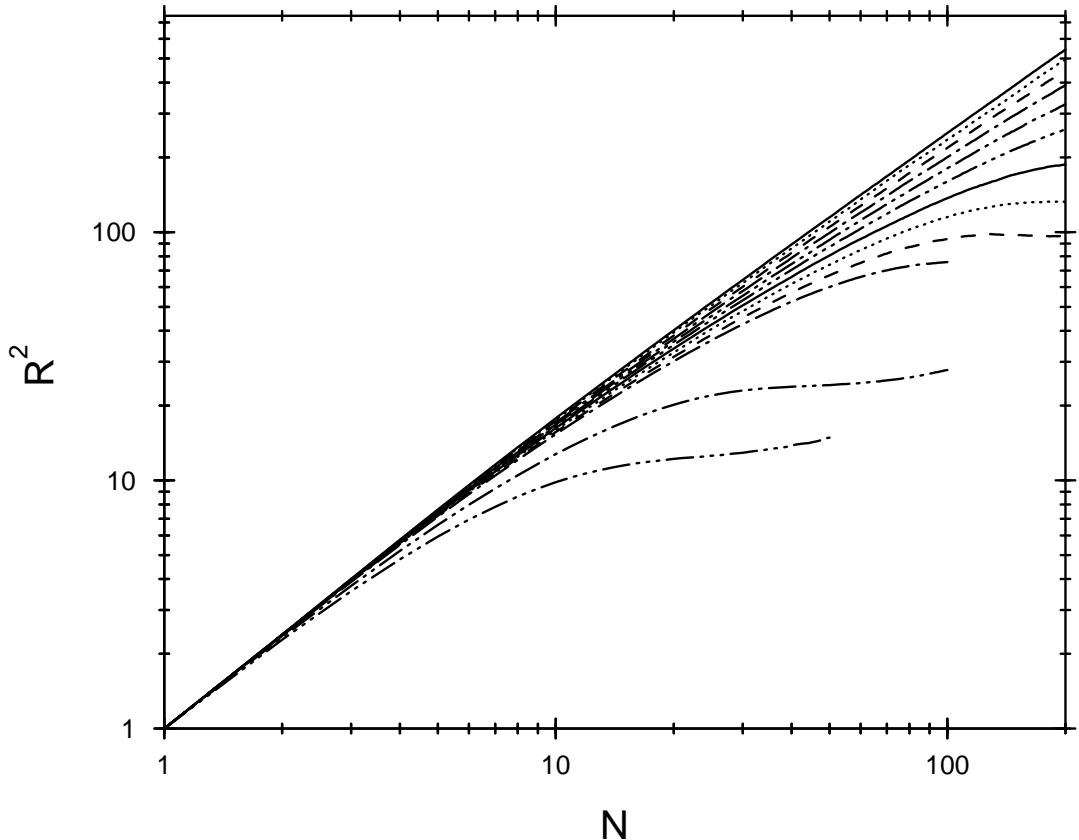}}
\caption{\small Simulation results for the average squared
end--to--end distance for $\beta=0.175,\ldots,0.375$ with
$\Delta\beta=0.025$ and additionally for $\beta=0.4,0.6,1/1.2$. 
The data clearly
show that the chains are already collapsed at $\beta\ge 0.25$.}
\label{fig.r2}
\end{figure}

\begin{figure}[htb]
\centerline{\epsfbox{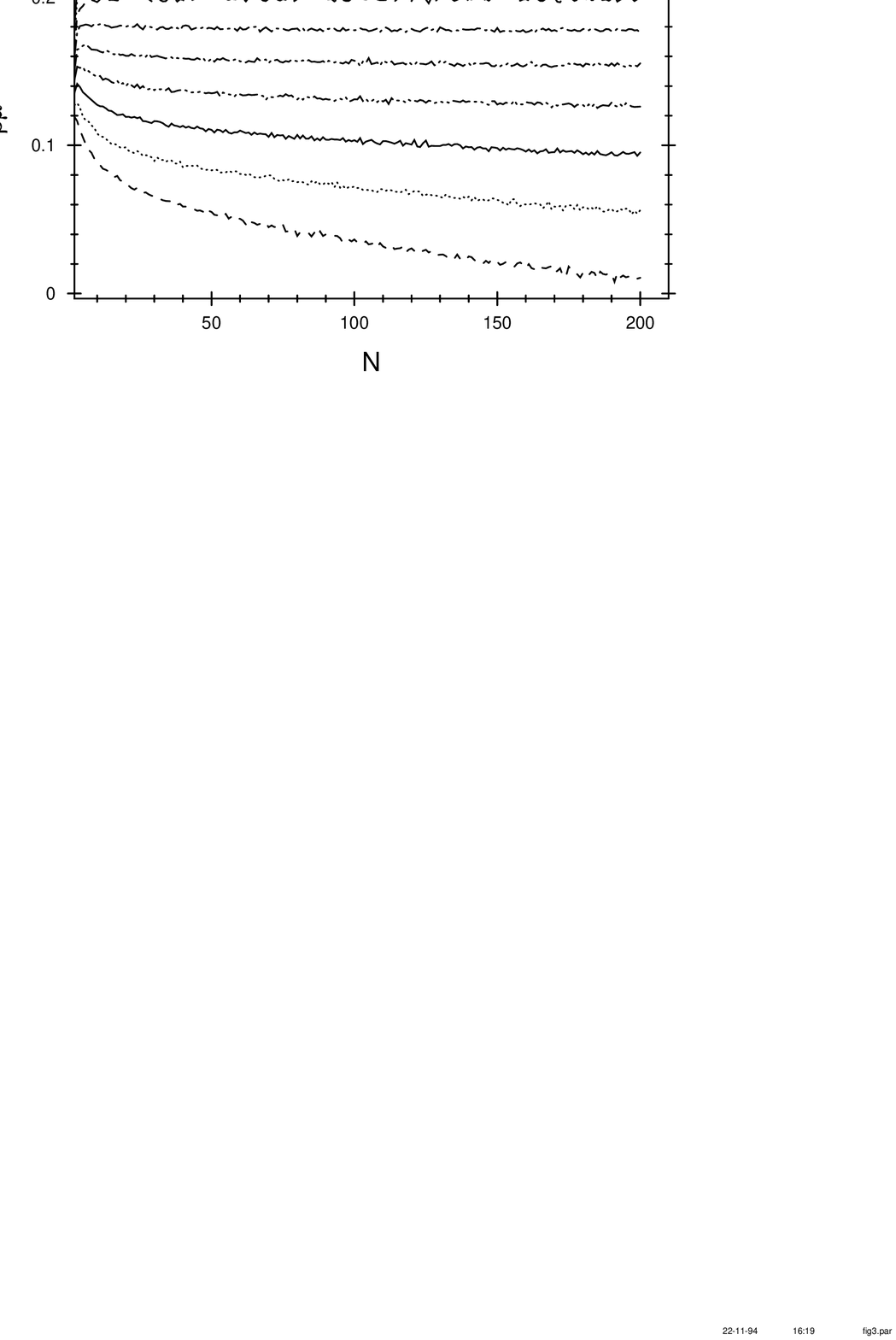}}
\caption{\small Simulation results for $\beta\mu^{\mathrm{inc}}$ for
$\beta=0.175,\ldots,0.375$ with $\Delta\beta=0.025$. One can see that
$\mu^{\mathrm{inc}}$ is independent of $N$ for $\beta\approx0.2$ to
$0.225$. So we can expect the $\Theta$--point somewhere between these
two values.}
\label{fig.dmudn}
\end{figure}

Averages of the squared end--to--end distance for longer chains are
shown in fig.~\ref{fig.r2}. They clearly indicate that $\beta=1/1.2$
is far below the $\Theta$--temperature in agreement with the fact that
$\mu^{\mathrm{ex}}$ is negative and decreasing with $N$. In
fig.~\ref{fig.dmudn} we show our data for
$\beta\mu^{\mathrm{inc}}=\beta(\mu_N^{\mathrm{ex}}-\mu_{N-1}^{\mathrm{ex}})$,
which represents the free energy needed to add one more monomer to the
chain. At the $\Theta$--point we expect $\mu^{\mathrm{inc}}$ to be
independent on $N$. The chains are still too short to pin down
$\beta_\Theta$ exactly, but the plots unambiguously show that
$\beta_\Theta$ is much smaller then $0.36$, the value given
in~\cite{kumar}. Together with the data from fig.~\ref{fig.r2} we
would say that $0.2\le\beta_\Theta\le0.23$.  Both, the data of
fig.~\ref{fig.r2} and that of fig.~\ref{fig.dmudn} were produced using
the hard rod potential for bonded monomers and the full LJ potential
for nonbonded monomers, i.e.\ without performing any cutoff at large
$r_{i,j}$. Each data set is based on at least $10^6$ 'tours'. We call
a tour the set of all (correlated) chains produced by the same initial
subroutine call from the main routine.

\section{Generalizations and Outlook}

In the above applications, we did not truncate the potential at large
$r$. Thus, inserting a new monomer takes a time ${\cal O}(N)$. For
larger $N$ this is no longer tolerable. If the potential is truncated
in such a case, one should also use efficient data structures for
searching relevant neighbors~\cite{schreiber}. With this we can
achieve ${\cal O}(1)$ behavior.

Our algorithm as described above can become inefficient for two main 
reasons, but in both cases this can be cured by minor modifications: 
first of all, if the temperature is very low, the Boltzmann factors 
$e^{-\beta \Delta U_i}$ for single monomers can become very large. The 
efficiency of the method results from the fact that large Boltzmann 
factors for the entire chain are split into smaller factors for 
individual monomers. If the latter factors themselves become large and 
are spread over a large interval, 
then most random positions will either be immediately discarded since 
they have very small $w_i$, or they will have very large $w_i$ and lead 
thus to very many replicas and correspondingly to huge fluctuations. 
A case where this made our method quite unsuccessful is the 
random heteropolymer model of \cite{iori}.

The simplest way out of this dilemma consists in choosing $M>>1$. In this 
way we will have even more points which are essentially useless since they 
have very small $w_i$. But the regions in state space with high probability 
will be sampled more evenly, and this should be more important. Another 
popular remedy consists in making the bulk of the simulations at higher 
temperatures, and quenching down to the desired temperatures in regular 
time intervals \cite{velikson}. 

The other case where the above algorithm gives poor results is provided by 
systems in which originally favored configurations lead finally to dead 
ends, while originally unfavored configurations become more important as 
the growth of the system continues. Two specific examples are dense polymer 
systems in 2-$d$ \cite{siepmann} and polymers with long range repulsive 
forces \cite{joensson,irbaeck}. 

Assume we want to grow a long self avoiding walk in a finite 2-$d$ region 
such that it fills a large fraction of the area. As we place the first 
monomers, all configurations are equally likely. But those which leave 
large closed voids all are dead ends since the walk later on cannot 
penetrate into the voids. Obviously it would be much better in this 
case to bias the walks against large voids from the very beginning. 

The situation is similar for a polymer with repulsive Coulomb potentials. 
There, unless the force is very strong, the effect of the repulsion is not 
too strong for end monomers, and hence new monomers are 
added without a strong radial bias. But the stretching force on a monomer 
deep inside the chain is much bigger (since all forces essentially add up), 
and the configuration is much more stretched inside the chain. Thus, when 
an existing chain is to be prolongated, most of the existing configurations 
have to be discarded, in order to be replaced by stretched configurations 
which at first (when they are assembled) are very improbable.

As we said, a way out of this problem is to use biased walks. This means 
that the number of replicas is not strictly proportional to $w_i$ but is 
larger in those regions which we suspect to become more important later. 
Of course it means also that we have to replace eq.(\ref{ZZZ}) by a weighted 
sum. It is e.g. known that the Rosenbluth trick \cite{rosen} leads to a 
bias towards more compact SAW's. We found indeed that our method with 
Rosenbluth weights instead of uniform weights was more efficient in 
giving SAW's which fill a square with periodic boundary conditions, but
it leads to much larger statistical fluctuations at low density. We should 
point out that this possibility of biasing is {\it independent} of the 
choice of $U^{(0)}$, something which seems to have been missed in 
\cite{garel,higgs,velikson}. 

Finally, we tried our method also for non-polymeric systems. For instance, 
we simulated the 2-d Ising model with spins numbered in the same way as 
they would be e.g. stored in a FORTRAN array. Though the method worked 
decently, it could not compete either with cluster flip methods (due to 
their much more efficient moves) nor with conventional Monte Carlo schemes 
which can be made very efficient by vectorization and multi-spin coding.
We should mention that the possible application spin models, and to the 
Ising model in particular, was also pointed out in \cite{niel} in the 
context of the breadth-first approach.

\vspace{.5cm}

P.G. wants to thank A. Vulpiani for hospitality at the University of Rome 
where part of this work was done. He also wants to thank D. Frenkel for 
correspondence and to S. Carraciolo, E. Marinari and G. Parisi for very 
helpful discussions. We are indebted to H. Meirovitch and I. Chang for 
pointing out refs. \cite{garel,higgs,velikson}. Unfortunately, this 
happened only after 
we had used our method in several papers (and after the present paper was 
essentially finished). We apologize to the above authors for not having 
cited them in previous papers. This work was supported by the DFG, 
Sonderforschungsbereich 237.

\eject


\begin{thebibliography}{99}

\bibitem{kumar-sl} S.K. Kumar, I. Szleifer and A.Z. Panagiotopoulos, Phys. 
     Rev. Lett. {\bf 66}, 2935 (1991)
\bibitem{kumar} S.K. Kumar, J. Chem. Phys. {\bf 96}, 1490 (1992)
\bibitem{harris} J. Harris and S.A. Rice, J. Chem. Phys. {\bf 88}, 
     1298 (1988)
\bibitem{pablo} J.J. de Pablo, M. Laso and U.W. Suter, J. Chem. Phys. 
     {\bf 96}, 6157 (1992)
\bibitem{laso} M. Laso, J.J. de Pablo and U.W. Suter, J. Chem. Phys. 
     {\bf 97}, 2817 (1992)
\bibitem{siepmann} J.I. Siepmann and D. Frenkel, Mol. Phys. {\bf 75}, 59 
     (1991)
\bibitem{frenkel} D. Frenkel, G.C.A.M. Mooij and B. Smit, J. Phys. 
     {\bf C 3}, 3053 (1991)
\bibitem{sheng} Y.-J. Sheng, A.Z. Panagiotopoulos, S.K. Kumar and 
     I. Szleifer, Macromolecules {27}, 400 (1994)
\bibitem{smit} B. Smit, G.C.A.M. Mooij and D. Frenkel, Phys. Rev. Lett.
     {\bf 68}, 3657 (1992)
\bibitem{kumar2} S.K. Kumar, I. Szleifer and A.Z. Panagiotopoulos, Phys. 
     Rev. Lett. {\bf 68}, 3658 (1992)
\bibitem{gh} P. Grassberger and R. Hegger, preprints (1993, 1994)
\bibitem{joensson} B. J\"onsson, C. Peterson and B. S\"oderberg, Phys. Rev. 
    Lett. {\bf 71}, 376 (1993); preprint LU-TP 93-15 (1993)
\bibitem{irbaeck} A. Irbaeck, Lund Univ. preprint LU-TP 93-20 (1993)
\bibitem{wall} F.T. Wall and J.J. Erpenbeck, J. Chem. Phys. {\bf 30}, 634, 
     637 (1959)
\bibitem{grishman} R. Grishman, J. Chem. Phys. {\bf 58}, 220 (1973) 
\bibitem{garel} T. Garel and H. Orland, J. Phys. {\bf A23}, L621 (1990) 
\bibitem{higgs} P.G. Higgs and H. Orland, J. Chem. Phys. {\bf 95}, 4506 
    (1991) 
\bibitem{velikson} B. Velikson, T. Garel, J.C. Niel and H. Orland, J. 
    Comp. Chem. {\bf 13}, 1216 (1992)
\bibitem{tarjan} R. Tarjan, SIAM J. Comput. {\bf 1}, 146 (1972)
\bibitem{gr-disorder} P. Grassberger, J. Phys. {\bf A 26}, 1023 (1993)
\bibitem{schreiber} T.\ Schreiber, to appear in Int.\ J.\ of Bif.\ and
    Chaos 
\bibitem{iori} G.Iori, E. Marinari and G. Parisi, J. Phys. {\bf A 24}, 
    5349 (1992) and preprint (1993)
\bibitem{rosen} M.N. Rosenbluth and A.W. Rosenbluth, J. Chem. Phys. 
    {\bf 23}, 356 (1955)
\bibitem{niel} J.C. Niel and H. Orland, preprint SPhT/90-137 (1990)

\end{thebibliography}
\end{document}